\documentstyle[11pt]{article}
\newcommand{\be}{\begin{equation}}
\newcommand{\ee}{\end{equation}}

\newcommand{\ts}{\hspace{3pt}}

\begin{document}
Physics Letters A {\bf 309}, 329--334 (2003)
\vspace*{.8cm}

\noindent
{\LARGE \bf  	There is no ``first'' quantization}

\vskip 1.5cm
\begin{quote}
\noindent
{\bf H. D. Zeh}
\vskip 0.2cm
\noindent
Universit\"at Heidelberg\\
www.zeh-hd.de
\end{quote}
\vskip 1.5cm

{\bf Abstract}
The introduction of spinor and other massive fields
by ``quantizing'' particles (corpuscles) is conceptually misleading.
Only spatial fields must be postulated to form the fundamental
objects to be quantized (that is, to define a formal {\it basis} for
all quantum states), while apparent ``particles'' are a
mere consequence of decoherence. This conclusion is also supported by
the nature of gauge fields.




\vskip1.2cm

\section {Introduction}
Decoherence theory \cite{Giulini} now allows us, in an appropriate
sense, to {\it derive} classical concepts in terms of universal quantum
theory. These classical concepts include clicks of a detector or spots
on a plate, phenomena which represent outcomes of measurements, and
which {\it seem} to indicate the presence of particles.
Particle concepts are therefore usually presumed in a ``quantization''
procedure (which leads to the $N$-particle wave functions of quantum
mechanics), and for an interpretation in terms of probabilities for the
``occurrence of values'' for particle properties
(regarded as ``observables'').

Relativistic theories, on the other hand, require quantum {\it field}
theory, where single particle wave functions, together with classical
fields, are used as arguments of field functionals, that now represent
the general (pure) {\it quantum states}. The ``occupation number''
representation, resulting for free fields (coupled oscillators), then
explains boson numbers by the numbers of
nodes in the wave functions for the amplitudes of all field modes.
This definition of ``particle'' number may be extended {\it beyond} the
harmonic approximation. Particle permutations are thereby reduced to
{\it identity operations} -- a consequence that is much deeper than a
mere {\it indistinguishability}. In
particular, it would explain the ``new statistics'' required
for presumed quantum particles.

These arguments suggest to
abondon a primordial particle concept entirely, and to
replace it with fields only. While this is indeed what has always been
done in the {\it formalism} of quantum field theory, particle concepts
are still used in a fundamental way for its interpretation, for
example when applied to scattering events. In a universal quantum
field theory, spatial fields (rather than particle positions) do not
only form the fundamental ``configuration'' space on which the wave
function(al) is {\it defined} as a general superposition.
Time-dependent quantum states may also describe apparently
discontinuous ``events'' by means of a smooth but rapid process of
decoherence.

So I agree with a consequence recently drawn by Ulfbeck and Bohr
\cite{UB} that ``no event takes place in the source itself as a
precursor to the click in the counter'', while I disagree with their
interpretation that the wave function ``loses its significance'' as
soon as an event occurs ``completely beyond law'' in the counter. On
the contrary, this event can be dynamically described in terms of a
unitarily evolving (hence strongly entangled) universal wave function.

This conclusion must also affect the interpretation of the
Wigner function, which is presently in vogue in non-relativistic
quantum mechanics because of its (misleading) formal analogy to a
classical phase space distribution.

\section {General quantum systems}
In order to set the stage for this comment, let me first define what
I mean by a general (abstract) quantum system. This is the conceptual
framework that remains when all {\it specific} aspects, such as those
resulting from ``quantizing'' a certain classical system, are
eliminated.

The {\it kinematics} of an abstract quantum system is defined by means
of a ``basis'' of linearly independent states
$|i\rangle$, with $i = 1,2,\dots,D$, subject to the superposition
principle, which allows every state of the system to be written in the
form
\be
|\alpha\rangle = \sum c_i\, |i\rangle \quad .
\ee
This kinematical principle requires furthermore that {\it any} such
(normalizable) superposition represents a possible physical state. For
a certain system, the dimension
$D$ may  be finite, infinite, or the states $|i\rangle$ may even form
a non-countable set. Although one can never strictly decide
empirically whether $D$ is infinite or just very large,
this difference is essential for the mathematical formulation of
explicit models.

The {\it dynamics} of a quantum system by itself is assumed to be
described by a Schr\"odinger equation,
\be
i{\partial \over \partial t}\, |\alpha\rangle = H\, | \alpha\rangle
\quad ,
\ee
characterized by a hermitean matrix $h_{mn}$ in the basis (1). In
quantum theories which contain gravity, the dynamics may degenerate to
a static Wheeler-DeWitt equation, $H|\alpha\rangle = 0$.

On the one hand, this is very little, since there is no interpretation
of these abstract states yet. On the other one, it is quite a bit,
since the superposition principle is known to be very powerful, while
the Hamiltonian may describe an enormous dynamical structure
(dynamical locality, for example).

Fortunately, in general we have more.

\section {Interpretation through measurements}
Measurements are interactions of the system with an appropriate
device. We know that there are specific system states $|x\rangle$, say,
which cause the ``pointer'' of a certain device
to move into a position that depends on the
state
$|x\rangle$. For general states $|\alpha\rangle$, this happens with
Born probalility $|\langle x | \alpha \rangle |^2$. For this purpose,
an inner product has to be added to the kinematics defined in Sect.\ts
2. Since different pointer positions exclude each other, we have to
require
$\langle x | x' \rangle = 0$ for $x \neq x'$.

There are various ways to describe such measurements.

(a) Traditional (Bohr): The pointer is described in classical terms.
Its ``position'' may be the actual position of a spot on the
photographic plate, or, for a different device, the impulse on a
macroscopic (Brownian) particle that we can observe under the
microcsope. In the first case, we usually presume (more or less
tacitly) that the measurement interaction is local, such that the
``quantum object'' must have {\it been} at this position, too.
Similarly, in the second case, we presume momentum conservation in
order to conclude that the quantum object must have changed its
momentum correspondingly, or must have lost it in the case of
absorption. (Bohr emphasized that conservation laws are essential for
the Copenhagen interpretation.) So one concludes that the ``quantum
object'' exhibits {\it properties of a particle} (position or momentum)
when being measured, even though we are forced to conclude that it
cannot {\it possess} both properties at the same time or when not
being observed. However, position and momentum may be used to define
two different bases for the corresponding quantum states, with
coefficients
$c_i$ becoming wave functions $\psi({\bf r})$ in the position
representation, where (1) assumes the form $|\alpha\rangle = \int
d^3r\ts
\psi({\bf r}) |{\bf r}\rangle$.

\looseness-1
The particle concept had proven useful earlier -- though not with
perfect results -- in statistical mechanics (for molecules) and in
Bohr-Sommerfeld quantum mechanics (for atomic electrons). By means of
formal considerations (based on the Hamiltonian form of mechanics)
this historical root led to a general ``quantization'' procedure,
applicable to classical dynamical systems. These quantization rules
and their consequences form the subject of this comment. For a
dynamical system that can be brought into Hamiltonian form, any
configuration space defines a basis for all quantum states, while
canonical momenta form another one (usually related to the former by a
Fourier transform -- only at this point an operator algebra based on
classical Poisson brackets becomes relevant). In this way one obtains
wave functions on configuration space, and, in particular, the well
established non-local many-particle wave functions
$\psi({\bf r}_1,\dots,{\bf r}_N,t)$. Note, however, that the concepts
of spin and permutation symmetry were {\it added} for empirically
reasons.

(b) Quantum pointers (von Neumann \cite{vN}):
Because of the generality of quantization rules, it appeared natural to
describe the pointer position ``x'' by a quantum state,
$|P_x\rangle$, too. If understood as a narrow wave packet of a massive
pointer, it may {\it approximately} define both position {\it and}
momentum (in accordance with the uncertainty relations or Fourier
theorem). For the specific states
$|x\rangle$, a measurement can then be written as a unitary evolution
in the tensor product space,
\be
|x\rangle \, |P_0\rangle \to |x\rangle\, |P_x \rangle \quad .
\ee
Apparently, Bohr was never ready to
accept this extension of the application of quantum theory to
macroscopic objects, even though he applied the uncertainty relations
to them. Equation (3) defines an effective interaction Hamiltonian
between system and pointer (neglecting all details), but inevitably
leads into the well-known measurement problem, which seems to require
either the existence of macroscopic superpositions (Schr\"odinger
cats) or a ``second dynamics'' (the collapse of the wave function).

(c) Universal quantum theory (Everett \cite{Everett}):
In the next ``natural step'', quantum theory was not only applied to
the system and its measurement device, but also to their environment
(the rest of the universe). Quantitative dynamical considerations then
require that all systems in the universe are strongly entangled
\cite{Z70}. Subsystems can possess quantum states by themselves only
``relative'' to states of the rest, in particular relative to states
of measurement devices or observers. The relation to observers is not
merely formal: it implies a radically novel definition of ``separate
observers'' in terms of the wave function -- required as a consequence
of entanglement. These relative states are factor states in dynamically
autonomous ``branches'' of the global wave function. Restricting
consideration to subsystems of the universe (as is realistic for local
interactions affecting local observers)
leads to the concept of decoherence and the formation of {\it
effective ensembles} of subsystem states
\cite{Giulini}. Phase relations defining (macroscopic) Schr\"odinger
cats are almost immediately dislocalized, and thus become irrelevant to
local observers. Since the universe is closed, there are no external
measurement devices to be used for an operational interpretation of
global (Everett) quantum states, and all observable properties must now
in principle be derived from the invariant structure of the universal
Hamiltonian.

\section {Quantization}
The ``canonical quantization rules'' require that classical
configuration variables $q$ define a basis of quantum states,
$|q\rangle$, for the ``corresponding'' Hilbert space.
Similarly, the quantum Hamiltonian is obtained from the classical
Hamiltonian $H(p,q)$ by replacing the canonical variables $p$ and $q$
with operators $P = \int dp\, |p\rangle p \langle p | $ and $Q = \int
dq\, |q\rangle q \langle q |$, respectively, (a procedure that cannot
be unique because of the factor ordering problem). We will here
essentially be concerned with the construction of the basis only,
which is then often used for a fundamental probability interpretation,
while the operators are defined to {\it act} on the quantum states
spanned by the basis in the form (1).

We now understand in principle how classical properties {\it
emerge} from the quantum system by means of decoherence (for example
as narrow Gauss packets in the canonical basis
$|q\rangle$): their superpositions would immediately decohere.
However, since decoherence depends on the environment, the question
arises whether the basis obtained by quantizing a classical theory is
always a fundamental one for the quantum system of interest.
If effective classical variables, $q(t)$, are known for a certain
system, we have to conclude according to the superposition principle
that all their superpositions $\int dq\, \psi(q,t)\, |q\rangle$ must in
principle exist as physical states, but this formal ``quantization''
procedure, based on classical concepts, does often {\it not} lead to a
{\it fundamental} basis (understood in a hierarchical sense if
required). Let me give three examples:

(1) The {\it rigid rotator} is classically described by means of the
Euler angles
$\phi,\theta,\chi$, say. Their symplectic structure defines the
geometry of this configuration space. Canonical quantization then leads
to wave functions
$D(\phi,\theta,\chi)$. They represent an effective approximation for
certain states of a many-body system (forming a rotational band),
which have more fundamentally to be described by a wave function
$\psi({\bf r}_1,\dots,{\bf r}_N)$. The stability of a rigid body,
required for this approximation, is itself based on {\it
quantum}  properties of the many-body system. In general, no
eigenstates for the Euler angles exist in terms of the
many-body states that form the rotational band
\cite{Z67}, since the approximation of a rigid body breaks down at
high values of angular momentum.

(2) {\it N-particle systems} would upon quantization lead to
wave functions depending on $3N$ position variables not restricted by
permutation symmetries. However, states of zero-mass bosons, for
example, can also be derived by quantizing a lattice or a
continuum of coupled oscillators (a ``field''). Its quantization leads
to equidistant energy levels (oscillator quanta), which can be
interpreted as boson numbers. It is for this reason that field
amplitudes appear as boson creation and annihilation operators. Since
{\it photon} number eigenstates are not robust against
decoherence,\footnote{  There is a popular misundstanding of
decoherence, found particularly in the context of {\it welcher Weg}
experiments. It assumes that decoherence is {\it defined} by the
disappearance of spatial interference fringes,  observable only in the
{\it statistics} of events. However, entanglement with an inaccessible
environment destroys phase relations between coefficients $c_i$ in
(local) {\it individual} quantum states (1). These superpositions
define {\it spatial} waves only in the special (though important) cases
of (effective) quantized single mass points or single oscillator quanta
on a spatial lattice. Apparent {\it events}, such as those appearing in
measurements and giving rise to statistical aspects, ``occur''
according to the Schr\"odinger equation in another (later) process of
decoherence. The latter affects superpositions of different
measurement outcomes, such as spots on a plate. In other situations,
depending on the relevant environment, {\it other}
(individual) quasi-classical states may be produced by decoherence, for
example precisely those {\it coherent states} of coupled oscillators
that define ``field'' modes
\cite{K+Z}. These two extremes of decoherence, caused by one or the
other measurement device, are conventionally interpreted as a ``wave
particle dualism''.}
their observed classical states are indeed {\it
fields}, which upon canonical quantization give rise to field
functionals rather than particle wave functions. For massive and, in
particular, charged boson fields (where ``mass'' is defined by a
specific term in the field equations), occupation numbers and
quasi-local states for each oscillator quantum are robust and may
therefore appear classical in many cases -- though not in the
non-environment-entangled state of their Bose-Einstein condensate
(cf.~\cite{Haensch}).   Permutation of bosons thus becomes a {\it
redundancy}, since this concept of bosons does not depend on a
primordial {\it particle} concept any more. For example, the
permutation symmetry of many-particle wave functions does {\it not}
represent any physical entanglement: it disappears in terms of wave
modes (cf. \cite{Shi}).

(3) The most important example (though not the subject of this Letter)
is the {\it dressing} of elementary fields. It may simply lead to a
``renormalization'' of parameters characterizing effective fields, or
even require quite new fields -- hopefully in the form of a
unification. While not restricted to {\it quantum} field theory (mass
renormalization was known in classical electron theory, for example),
the results of dressing must be expected to depend essentially on
quantum theory.

Fermions (but also massive bosons) are usually regarded as particles on
a fundamental level. This is particularly evident in Bohm's theory
\cite{Bohm}, where unobservable
trajectories in classical configuration space are postulated for
particles {\it and} for electromagnetic and other fields. (Variants of
Bohm's theory with photon
trajectories instead of time-dependent Maxwell fields have recently
been claimed to be in conflict with quantum theory and experiments
\cite{antiBohm}. While the analysis of these experiments appears
doubtful, this modified Bohm theory -- in contrast to the original one
-- seems to have never been proven equivalent to quantum theory.)

In the Heisenberg picture, ``observables'' corresponding
to classical particle variables are often assumed to ``exist'' but not
to ``possess values''. However, since many-particle {\it quantum}
states are represented by wave functions (on a space of $3N$
dimensions), while particle aspects, such as spots on a plate or
clicks in a counter, {\it emerge}  by means of decoherence in
accordance with a universal Schr\"odinger equation, I concluded ten
years ago
\cite{Z93} that ``there are no particles'' in quantum mechanics any
more. Their  r\^ole in the quantization procedure (for
defining the corresponding configuration space as a stage for the wave
function) is nonetheless widely used as an argument for a probability
interpretation in terms of particles.

In canonical quantum electrodynamics, wave functionals $\Psi[\psi
({\bf r}),{\bf A}({\bf r}),t]$ or $\Psi[\psi({\bf r}),\psi^\ast
({\bf r}),{\bf A}({\bf r}),t]$ describe general quantum states.
They represent entangled superpositions of different values of all
field amplitudes, thus leading to field {\it operators} and their
canonical momenta for
${\bf A}(\bf r)$ and $\psi(
\bf r)$. Since
$\psi({\bf r},t)$ was itself obtained from particle quantization,
this procedure is often called a ``second'' quantization.  This
interpretation is obviously wrong, since a true second quantization
would lead to wave functionals defined on {\it many}-particle wave
functions. The ``one-particle wave function'' $\psi({\bf r}, t)$ is a
perfectly local {\it field}, that would not allow one to describe EPR
type non-locality, for example. While a quantized spinor field was {\it
historically} a second step, we must now simply conclude that spinor
fields (rather than particle positions) define a correct {\it basis}
for electron and other fermion quantum states, even though they hardly
ever appear as quasi-classical objects. Position
${\bf r}$ never represents a dynamical variable; it occurs as an {\it
index} of the true variables ($\psi$ and
  other dynamical fields).

In other words: there are not even particles ``before''
quantization, that is, characterizing in any way
the ``configuration'' space on which a fundamental wave function(al) is
defined.  According to the present state of the art, the ``second''
quantization in terms of fields is the first and only one,
while particles represent a derived and {\it effective} concept. Their
appearance is no more than the result of decoherence by means of local
interactions: it leads to robust (quasi-classical) local effects
in the cloud chamber or detector, representing droplets or clicks,
respectively. The occupation number (rather than particle) basis for
electron states has recently been experimentally confirmed by
anti-bunching
\cite{Hassel}, while the non-invariance of neutrons under
$2\pi$-rotations was directly observed long ago
\cite{Rauch}. This double-valuedness of spinor fields under full
spatial rotations may explain the restriction of their
occupation numbers to 0 and 1.

If a modified Bohm theory with photon trajectories is indeed in
conflict with quantum theory, I would expect this
to apply to Bohm's original theory in the {\it relativistic}
case, too. One may instead need Bohm trajectories for
fields only \cite{Samols} in order to remain consistent
with relativistic quantum theory and with experiments (while
even these ``consistent'' Bohm trajectories remain unobservable
and in this sense meaningless
\cite{Z99}).

The conclusion that there is only a quantum theory of {\it fields}
does not, of course, contain any novel consequences for the {\it
formalism} of conventional quantum field theory. However, it
undermines  the usual interpretation of quantum states as
probability amplitudes for (conceptually primordial) particles. This
consequence may affect also other aspects of the Heisenberg picture.
One may even argue whether a functional of {\it fields} will survive in
a {\it future} quantum theory. Its form as a functional of
many effective ``particle'' fields is certainly the most successful
theory yet, but unified quantum field theories (supersymmetry or
M-theory, for example) are no more than promising {\it proposals} for
a more fundamental one.  Quantum field theories have the important
advantage, however, to allow the formulation of {\it local} dynamics by
means of a Hamiltonian density, defined as a function of field
operators and their derivatives.

\section {Constraints}
Gauge theories are using constraints, which may be understood as a
means to eliminate ``unphysical'' degrees of freedom, or redundancies
(``gauges''), in order to define the proper {\it stage} for the wave
function. For example, the permutated positions of two ``identical
particles'', or two magnetic potentials which lead to the same
magnetic field and loop integrals
$\int {\bf A}({\bf r})\cdot d{\bf s}$, have to be physically
identified. While this {\it should} be done before quantization, that
is, when defining the Hilbert space basis in (1), it is
often more convenient, or the only feasible way, to apply ``quantum
constraints'' in the form
\be
C\, |\alpha\rangle = 0
\ee
to an unphysical (too
large) Hilbert space based on unconstrained variables. The latter is
thus restricted to states being symmetric (invariant)  under the group
of all unphysical transformations, such as $\exp(iC\phi )$
(see \cite{GKZ} and Giulini's Sect.~6.3 of
\cite{Giulini} for a relation to superselection rules).   These two
procedures are expected to be equivalent, while the enlarged Hilbert
space remains irrelevant for any physical interpretation.

In quantum gravity, for example, ``momentum constraints'' $P_i
|\alpha \rangle = 0$, applied to the Hilbert space spanned by all
spatial metrics
$h_{kl}({\bf r})$, with $i,k,l=1,2,3$, are known to symmetrize
the physical states under all transformations which connect
different spatial metrics that represent the {\it same} abstract
  spatial geometry (such as those related by a mere coordinate
transformation). A coordinate-free description of three-geometry is
not explicitly known, in general.

There is a catch, however. Invariance under the gauge proper  (the
local choice of a basis of gauge group generators) is related
to the existence of a {\it physical} ``gauge field'' (see the {\it
note added in proof} in
\cite{YM}). The parallel transport of these generators (a connection
on the corresponding fibre bundle), which is required for the
meaningful definition of {\it relative} gauge
transformations at different points, must itself represent
gauge-independent (abstract) {\it geometry}, and may thus give
rise to {\it active} (physical) oscillations (``bosons'' after their
quantization). It is for this physical reason that the gauge field
``should be varied in the Lagrangean''
\cite{YM}. Invariance then holds trivially under local basis
transformations, but (if defined) also under
active {\it global} ones (Mach's principle understood as a redundancy).

This picture supports the view, entertained
in this comment, that even fermions have to be
fundamentally described by quantum {\it fields} rather than
quantum particles. These fields may carry properties (``charges'') that
are affected by transformations under the gauge group -- for example a
``classical'' phase characterizing a complex spinor field. This phase
does {\it not} describe a quantum superposition (as it would for
quantized charged {\it particles}). In Weyl's classical gauge
theories, their later application (in 1929) to the ``quantum'' phase
(see
\cite{Strau}) appears as a {\it deus ex machina}, while it would be
entirely natural for complex pre-quantum (``classical'') spinor {\it
fields}.

In the same sense as the observation of a radiation reaction in the
absence of any absorbers \cite{RR} has been regarded, in classical
context, as evidence for the {\it reality of fields} (in contrast to
time-symmetric action at a distance), decoherence of the
source by its own radiation in the absence of any events in
absorbers would demonstrate the {\it reality of the corresponding wave
functional} (which here describes entanglement between the radiation
and its source).
\medskip

\noindent {\bf Acknowledgement:} I wish to thank the participants of
the FESt seminar on decoherence (in particular Domenico Giulini and
Claus Kiefer), Shelly Goldstein and Peter Holland for helpful
discussions or comments.

\end{document}